%% file: main.tex
\documentclass[journal,twocolumn]{IEEEtran}
\usepackage[obeyFinal]{easy-todo} % todo notes
\usepackage{soul} % Strikethrough text using \st{}
\usepackage{cite}             % improves presentation of citations
\usepackage{amssymb} % for math symbols like ~
\usepackage{amsmath}
\usepackage{amsthm} % for \newtheoremstyle
\usepackage{aligned-overset}
\usepackage[bookmarks=false]{hyperref}
\newtheoremstyle{mytheorem}% name of the style to be used
  {0pt}% measure of space to leave above the theorem. E.g.: 3pt
  {0pt}% measure of space to leave below the theorem. E.g.: 3pt
  {\itshape}% name of font to use in the body of the theorem
  {}% measure of space to indent
  {\bfseries}% name of head font
  {:}% punctuation between head and body
  {0.5em}% space after theorem head; " " = normal interword space
  {}% Manually specify head

\theoremstyle{mytheorem}
\newtheorem{theorem}{Theorem}
\newtheorem{lemma}{Lemma}

\newtheorem{example}{Example}

\newtheorem{definition}{Definition}

\newcommand{\Brack}[1]{\left({#1}\right)} %Brackets
\newcommand{\floor}[1]{\left \lfloor {#1} \right \rfloor }

\newcommand{\norm}[1]{\left\Vert#1\right\Vert}

\newcommand{\bb}{\mathbf{b}}

\newcommand{\bx}{\mathbf{x}}
\newcommand{\cV}{{\cal V}}
\newcommand{\cC}{{\cal C}}
\newcommand{\cS}{{\cal S}}
\def\b{\mathbf}
\NewDocumentCommand\sqn{mg}{%
    \|\mathbf{#1}_{\IfNoValueTF{#2}{}{#2}}\|^2%
}

\DeclareMathOperator{\EX}{\mathbb{E}}% expected value
\usepackage{graphicx}
\graphicspath{{figs/}}
\usepackage{subcaption}

\usepackage{tikz}
\usetikzlibrary{decorations.pathreplacing,calligraphy}
\usepackage{subfiles} % Best loaded last in the preamble
\begin{document}

\title{Correction to "Private Information Retrieval Over Gaussian MAC"}

\author{\IEEEauthorblockN{Or Elimelech, Ori Shmuel and Asaf Cohen}}
\maketitle
%%%%%%%%%%%%%%%%%
%%%%%%%%%%%%%%%%%
\input{abstract}
\begin{IEEEkeywords}
Private Information Retrieval, Multiple Access Channel, Lattice Codes.
\end{IEEEkeywords}
%%%%%%%%%%%%%%%%%
%%%%%%%%%%%%%%%%%
% \input{Intorduction}
%%%%%%%%%%%%%%%%%
%%%%%%%%%%%%%%%%%
\input{Joint_NonFading}
%%%%%%%%%%%%%%%%%
%%%%%%%%%%%%%%%%%
\input{Joint_Fading}
%%%%%%%%%%%%%%%%%
%%%%%%%%%%%%%%%%%
% \input{Conclusion}

%%%%%%%%%%%%%%%%%
%%%%%%%%%%%%%%%%%
\input{appendices}

\begin{IEEEbiographynophoto}{Or Elimelech}
received his B.Sc. degree in Electrical Engineering and his M.Sc. degree in Communication Systems Engineering from Ben-Gurion University of the Negev, Be’er Sheva, Israel, in 2021 and 2023, respectively. He is currently pursuing a Ph.D. in the School of Electrical and Computer Engineering at the same institution. His research focuses on information theory and physical layer security, with recent works exploring privacy and secrecy in modern communication networks.
\end{IEEEbiographynophoto}

\begin{IEEEbiographynophoto}{Ori Shmuel}
received his B.Sc. (summa cum laude), M.Sc. (cum laude), and Ph.D. degrees from the Department of Communication Systems Engineering at Ben-Gurion University of the Negev, Be’er Sheva, Israel, in 2013, 2015, and 2021, respectively. He is currently a Principal System Architect for 5G communication products with Parallel Wireless Inc., where his primary focus is on MaMIMO technology in the context of 5G and AI applications for wireless communication. His main research interests include wireless communication, distributed systems, user scheduling, information theory, and ML\&AI.
\end{IEEEbiographynophoto}

\begin{IEEEbiographynophoto}{Asaf Cohen}
Asaf Cohen received the B.Sc. (Hons.), M.Sc. (Hons.), and Ph.D. degrees from the Department of Electrical Engineering, Technion, Israel Institute of Technology, in 2001, 2003, and 2007, respectively. From 1998 to 2000, he was with the IBM Research Laboratory, Haifa, where he was working on distributed computing. Between 2007 and 2009 he was a Post-Doctoral Scholar at the California Institute of Technology, and between 2015–2016 he was a visiting scientist at the Massachusetts Institute of Technology. He is currently an Associate Professor and the Vice Chair for Teaching at the School of Electrical and Computer Engineering, Ben-Gurion University of the Negev, Israel. His areas of interest are information theory, learning, and coding. In particular, he is interested in network information theory, network coding and coding in general, network security and anomaly detection, statistical signal processing with applications to detection and estimation and sequential decision-making. He received several honors and awards, including the Viterbi Post-Doctoral Scholarship, the Dr. Philip Marlin Prize for Computer Engineering in 2000, the Student Paper Award from IEEE Israel in 2006 and the Ben-Gurion University Excellence in Teaching award in 2014. He served as a Technical Program Committee for ISIT, ITW and VTC for several years, and as an Associate Editor for Network Information Theory and Network Coding; Physical Layer Security; Source/Channel Coding and Cross-Layer Design to the IEEE Transactions on Communications.
\end{IEEEbiographynophoto}

%%%%%%%%%%%%%%%%%
%%%%%%%%%%%%%%%%%
\bibliographystyle{IEEEtran}
\bibliography{references}
\end{document}

%% file: Abstract.tex
\begin{abstract}
In the above article \cite{shmuel2021private}, the authors introduced a PIR scheme for the Additive White Gaussian Noise (AWGN) Multiple Access Channel (MAC), both with and without fading.
The authors utilized the additive nature of the channel and leveraged the linear properties and structure of lattice codes to retrieve the desired message without the servers acquiring any knowledge about the retrieved message's index.

Theorems 3 and 4 in \cite{shmuel2021private} contain an error arising from the incorrect usage of the modulo operator. 
Moreover, the proofs assume a one-to-one mapping function, $\phi(\cdot)$, between a message $W_j\in\mathbb{F}_p^L$ and the elements of $\cC$, mistakenly suggesting that the user possesses all the required information in advance.
% \st{However, this is not the case.}
\textcolor{black}{To deal with that, we defined $\phi(\cdot)$ as a one-to-one mapping function between a vector of $l$ information bits and a lattice point $\lambda\in\cC$}. 
Herein, we present the corrected versions of these theorems.
\end{abstract}

%% file: Joint_NonFading.tex
\section{Corrected Theorems}
A mistake in the analysis of the decoding stage in \cite{shmuel2021private} led to an error in \cite[Theorem 3]{shmuel2021private} and \cite[Theorem 4]{shmuel2021private}.
Please refer to Appendix \ref{Appendix:ErrorExplain} for a detailed explanation.
\textcolor{black}{
The corrected schemes, shown in the proof, have two main changes. 
The first is in the query vector. 
In the previous scheme, the user generates a random vector $\mathbf{b}$ of length $M$ such that each entry is either $1$ or $-1$.
In the new scheme, the user generates a random vector $\mathbf{b}$ of length $M$ such that each entry is $1$ or $0$.
This change helped overcome the mathematical issues in the previous scheme. 
We note that this change resulted in only half of the server's answers containing the desired message, which degraded the rate by $1$ bit per channel use. This can be seen more intuitively in the non-fading case (Theorem \ref{theorem:NonFading}), with the $\left\lfloor \frac{N}{2} \right\rfloor^2$ coefficient. 
The second change is more fundamental.
In the previous scheme, there was a one-to-one mapping
function, $\phi(\cdot)$, between a message $W_j\in\mathbb{F}_p^L$ and the elements of $\cC$, mistakenly suggesting that the user possesses all the required information in advance.
To correct that, we redefined $\phi(\cdot)$ as a one-to-one mapping function between a vector of $l$ information bits and a lattice point $\lambda\in\cC$. 
Now, each lattice point $\lambda\in\cC$ represents $l$ bits string instead of one of the servers' messages.}

This section presents a corrected version of the theorems.
\begin{theorem}\label{theorem:NonFading} For an AWGN MAC with $N$ servers, the achievable PIR rate is given by: 
\begin{equation}
    R^{J}_{PIR}=\frac{1}{2} \log^+\left(\frac{1}{2}+\left\lfloor \frac{N}{2}\right\rfloor^2P\right)
\end{equation}
\end{theorem}

Note that the main difference from the original theorem is that $\left\lfloor \frac{N}{2} \right\rfloor^2$ is not multiplied by $4$. Consequently, at most $1$ bit is lost per channel use.

Please refer to Appendix \ref{Appendix:NonFadingPIRScheme} for a self-contained corrected proof. 
The correction in the theorem above is achieved through modifying the scheme.
% \st{In the corrected version, only half of the server's answers contain the desired message, while the others are utilized to cancel out unwanted messages.}

As a result, the gap between the Multiple-Input Single-Output (MISO) sum capacity and the PIR rate presented in \cite[Lemma 1]{shmuel2021private} should also be updated.
\textcolor{black}{As far as we know, the capacity for the PIR over AWGN MAC remains unsolved. Hence, we use the MISO capacity as an upper bound for our problem.}

\begin{lemma}\label{lemma:gap}
The PIR rate for the AWGN MAC with $N$ servers, as given in Theorem \ref{theorem:NonFading}, has a finite gap from the channel capacity, denoted as $C^{MISO}_{SR}$. Specifically:

\begin{equation*}
    C^{MISO}_{SR}-R^J_{PIR}\leq  
    \begin{cases}
     1 \ , \qquad N \ \text{mod} \ 2=0
     \\
     2 \ , \qquad N \ \text{mod} \ 2=1
    \end{cases}  
\end{equation*}

\end{lemma}

Please refer to Appendix \ref{Appendix:gapLemmaProof} for the proof. 
The result above highlights that privacy is achieved at the cost of a minor reduction in the rate for even values of $N$, and an additional negligible loss for odd values of $N$.

%% file: Joint_Fading.tex
Due to the same error, Theorem 4 should be corrected as well.
\begin{theorem}\label{theorem:Fading}
Consider an $N$ servers, block-fading AWGN MAC. 
For any non-empty subsets of servers $S_1, S_2$ satisfying ${\mathcal{S}_1\cap \mathcal{S}_2=\emptyset,\ \mathcal{S}_1 \cup \mathcal{S}_2 \subseteq\{1, ..., N\}}$ 
and any integer vector $\mathbf{a}\in\mathbb{Z}^2$ with non-zero entries, the following PIR rate is
achievable,
\begin{equation}\label{eq:fading_rate}
R^J_{PIR}=\frac{1}{2}\log^{+}\Brack{\frac{1+P\tilde{\norm{\mathbf{h}}}^2}{\norm{\mathbf{a}}^2+P(a_1\Tilde{h_2}-a_2\Tilde{h_1})^2}}
\end{equation}
where, $\mathbf{\tilde{h}}=(\Tilde{h}_1,\Tilde{h}_2)\in \mathbb{R}^2 \ and \ \Tilde{h}_1=\sum_{k\in \mathcal{S}_1} {h_k}$ and $\Tilde{h}_2=\sum_{k\in \mathcal{S}_2} h_k$.\\
\end{theorem}
It is worth noting that, once again, the main difference from the original theorem is that the numerator is not ${\text{multiplied by } 4}$.
A self-contained, corrected proof is in Appendix \ref{Appendix:FadingPIRScheme}.

%% file: appendices.tex
\appendices
%%%%%%%%%%%%%%%%%%
%%%%%%%%%%%%%%%%%%
\section{Error Explained}\label{Appendix:ErrorExplain}
We first mention some definitions and key properties of nested lattices.
The notation is similar to \cite{shmuel2021private,nazer2011compute}.
\begin{definition}[Quantizer]
A lattice quantizer is a map, ${Q_{\Lambda}: \mathbb{R}^n \rightarrow  \Lambda}$, that sends
 a point, $\mathbf{s}$, to the nearest lattice point in Euclidean distance. That is,
 \begin{equation}
     Q_{\Lambda}(\mathbf{s})=\text{argmin}_{\lambda\in\Lambda}\norm{\mathbf{s-\lambda}}.
 \end{equation}
\end{definition}
\begin{definition}[Voronoi Region]
The \textit{fundamental Voronoi region}, $\mathcal{V}$, of a lattice, $\Lambda$, is the set of all points in $\mathbb{R}^n$ that are closest to the zero vector compared to any other lattice point. That is, 
${\mathcal{V}=\{\mathbf{s}: Q_{\Lambda}(\mathbf{s})=0\}}$.
\end{definition}
\begin{definition}[Modulus \textcolor{black}{as Quantization Noise}] Let $[\mathbf{s}] \text{mod} \ \Lambda$ denote the quantization error of $\mathbf{s}\in\mathbb{R}^n$ with respect to the lattice $\Lambda$. That is,
\begin{equation}
[\mathbf{s}] \text{mod} \ \Lambda =\mathbf{s}-Q_{\Lambda}(\mathbf{s}).
\end{equation}   

For all $\mathbf{s,t} \in \mathbb{R}^n$ and $\Lambda_c \subseteq \Lambda_f$, the $\text{mod} \ \Lambda$ operation satisfies:
\begin{equation}\label{eq:dist}
[\mathbf{s} + \mathbf{t}] \text{mod} \ \Lambda =
\big[[\mathbf{s}] \text{mod} \ \Lambda +\mathbf{t}\big] \text{mod} \ \Lambda 
\end{equation}

\begin{equation}
\left[Q_{\Lambda_f}(\mathbf{s})\right]\ \text{mod} \ \Lambda_c =
\left[Q_{\Lambda_f}([\mathbf{s}]\ \text{mod} \ \Lambda_c)\right]\ \text{mod} \ \Lambda_c 
\end{equation}

\begin{equation}\label{eq:Zscal}
[a\mathbf{s}] \text{mod} \ \Lambda =
[a[\mathbf{s}] \text{mod} \ \Lambda]\text{mod} \ \Lambda \quad \forall a\in \mathbb{Z} 
\end{equation}

\begin{equation}\label{eq:scal}
\beta [\mathbf{s}] \text{mod} \ \Lambda =
[\beta \mathbf{s}] \text{mod} \ \beta \Lambda \quad \forall \beta\in \mathbb{R} 
\end{equation}

\end{definition}
%%%%%%%%%%%%%%
\subsection{Errors}
On Page 9 in \cite{shmuel2021private}, there are two errors that need to be addressed. The first is in the fourth equality (\emph{a}), and the second is in the fifth equality (\emph{b}).
The first error arises from the following equality:
\begin{equation}\label{eq:problem1}
\begin{split}
    \Big[\alpha \big([\mathbf{A}_{1}-\mathbf{d}]\text{mod} \ \Lambda_c +[\mathbf{A}_{2}-\mathbf{d}]\text{mod} \ \Lambda_c\big)\Big]\text{mod} \ \Lambda_c
    \\ = 
    \Big[\alpha \big([\mathbf{A}_{1}+\mathbf{A}_{2}]\text{mod} \ \Lambda_c -[2\mathbf{d}]\text{mod} \ \Lambda_c\big)\Big]\text{mod} \ \Lambda_c    
\end{split}
\end{equation}
where $\mathbf{A}_{i}\in \Lambda_f$, $d\in\mathbb{R}^n$ and $\alpha \in \mathbb{R}$.
% \st{It should be noted that the expression on the left-hand side is not always equal to that on the right-hand side.}
\textcolor{black}{The equality is true only when $\alpha \in \mathbb{Z}$.} 
In our case, this equality is incorrect since $\alpha \in \mathbb{R}$. 
% \st{is not necessarily an integer;} Therefore, property \eqref{eq:Zscal} cannot be used. 
To illustrate this, consider the following example:

\subfile{figs/LatticeExample}

\begin{example}
Assume a one dimension Nested-Lattice where $\Lambda_f=\mathbb{Z}$, $\Lambda_c=5\mathbb{Z}$,
i.e., $\Lambda_c\subseteq\Lambda_f$ (Figure \ref{fig:NestedLattice}).
Let $\alpha=\frac{1}{2}$, $\mathbf{A}_{1}=2$, $\mathbf{A}_{2}=1$, and $\mathbf{d}=0$.
For the left hand-side in (\ref{eq:problem1}) we have the following:
$$ \Big[\frac{1}{2} \big([2]\text{mod} \ \Lambda_c +[1]\text{mod} \ \Lambda_c\big)\Big]\text{mod} \ \Lambda_c=\frac{3}{2}$$
and for the right-hand-side, we have:
$$\Big[\frac{1}{2} \big([2+1]\text{mod} \ \Lambda_c\big)\Big]\text{mod} \ \Lambda_c 
= \Big[\frac{1}{2} \big(-2\big)\Big]\text{mod} \ \Lambda_c = -1 $$
As we see, the equality does not hold.
\end{example}
The second error contains at the fifth equality \emph{(b)} and equation (22) is as follows,
\begin{equation}\label{eq:problem2}
    \Big[\alpha\big[\mathbf{A}_{1}+\mathbf{A}_{2}\big]mod \ \Lambda_c\Big]mod \ \Lambda_c \neq
     [\alpha(2v_i)] \ mod \ \Lambda_c
\end{equation}
again this equality does not hold because $\alpha \in \mathbb{R}$.
%%%%%%%%%%%%%%%%%%
%%%%%%%%%%%%%%%%%%
\section{Corrected Proof for Theorem \ref{theorem:NonFading}}\label{Appendix:NonFadingPIRScheme}
\begin{IEEEproof}
The user, which is interested in message $W_i$, generates a random vector $\mathbf{b}$ of length $M$ such that each entry is either $1$ or $0$, independently and with equal probability.
Then, the user creates a second vector $\bb'$  by the following steps. 
First, take NOT on the \textcolor{black}{$i$th} bit, then apply minus on the vector.
% \st{Note that $B\thicksim U\{0,1\}$ and $B'\thicksim U\{-1,0\}$.}
Then the user sends the following queries to the $k$'th and $(k+1)'$th servers,
\begin{equation} \label{eq:queries}
    \begin{split}
       &Q_k(i)=\mathbf{b}, \ Q_{k+1}(i)=\mathbf{b'}=-\mathbf{b}-\mathbf{e}_i\text{, if $b_i = 0 $} \\
       &Q_k(i)=\mathbf{b}, \ Q_{k+1}(i)=\mathbf{b'}=-\mathbf{b}+\mathbf{e}_i\text{, if $b_i = 1 $} 
    \end{split}
\end{equation}
Where $k\in \{1,3,5...,2\floor{\frac{N}{2}}-1\}$.
If $N$ is odd, the $N$th server is ignored.
The above creates a partition of the servers into two equal-size groups with the same query for all group members.
From the servers' perspective, each sees a uniform random vector depending on their group.
The first group sees a uniform random vector with an element being $0$ or $1$ with equal probability. 
The other group sees a uniform random vector with an element being $-1$ or $0$ with equal probability. 
Thus, the privacy of the index $i$ is guaranteed.
Specifically, following the privacy requirement \cite[(2)]{shmuel2021private}, we have,
\begin{equation}\label{eq:privacy}
I(\theta;Q_j(\theta),\mathbf{x}_j(\theta),W_1^M)=0 \text{ for all }j\in\{1,...,N\}    
\end{equation}
the proof is the same as in \cite{shmuel2021private}.\\
We assume the messages are long. 
Hence, the message is broken up into smaller, manageable chunks of information (packetization) $\b{s}_{j,m}\in \{0,1\}^l$ where $\b{s}_{j,m}$ is the $j$th packet of $W_m$ \textcolor{black}{and $\mathbf{v}$ denotes the mapped lattice codeword for the requested packet $\mathbf{s}_{j,m}$, specifically,  $\mathbf{v} = \phi(\mathbf{s}_{j,m})$.}
These packets are then transmitted over the channel and reassembled by the receiver to form the original message.
Hence, we focus on the transmission of the $j$th packet.
The rest are transmitted the same way.

The servers use the Compute and Forward (CF) coding scheme \cite{nazer2011compute}, which uses nested lattice codebooks.
Specifically, we construct a nested lattice codebook as in \cite[Section 4.B]{nazer2011compute}, where $\Lambda_c$ and $\Lambda_f$ are a pair of $n$-dimensional lattices with Voronoi regions $\cV_c$ and $\cV_f$, respectively, such that $\Lambda_c$ is a subset of $\Lambda_f$, i.e., $\Lambda_c \subset \Lambda_f$.
That is, the nested lattice code is given by $\cC=\{\Lambda_f \cap \cV_c\}$.
The code is known to the user and all the servers.
In addition, there exists a one-to-one mapping function, $\phi(\cdot)$, between a vector of $l$ information bits and a lattice point $\b{\lambda}\in\cC$ \cite[Lemma 5]{nazer2011compute}, namely:
$$\b{s} = (s_1,...,s_l)\in\{0,1\}^l \mapsto \b{\lambda}=(\lambda_1,...,\lambda_n)\in\cC$$
Note that the codebook does not depend on the messages since it contains all combinations of words with length $l$ bits.

Upon receiving the queries, the servers form their answers by performing modulo-$\Lambda_c$ addition between the mapped packets (i.e., lattice codewords)  multiplied by their corresponding entry in $Q_j(i)$.
That is,

\begin{equation} \label{eq:answers}
    \begin{split}
      &\mathbf{A}_{k}=\left[\sum_{m=1}^{M}Q_{k,m}(i)\phi(\b{s}_{j,m})\right] \text{mod} \ \Lambda_c \\
      &\mathbf{A}_{k+1}=\left[\sum_{m=1}^{M}Q_{k+1,m}(i)\phi(\b{s}_{j,m})\right] \text{mod} \ \Lambda_c \\
    \end{split}
\end{equation}

where $k\in \{1,3,5...,2\floor{\frac{N}{2}}-1\}$; $Q_{k,m}(i)$ is the $m$th entry of the vector $Q_k(i)$;
Note that ${[\mathbf{A}_k + \mathbf{A}_{k+1}] \ \text{mod} \ \Lambda_c}$ is equal to either ${[\phi(\b{s}_{j,i})] \ \text{mod} \ \Lambda_c=[\b{v}] \ \text{mod} \  \Lambda_c}$,
or $[-\phi(\b{s}_{j,i})] \text{mod} \ \Lambda_c=[-\b{v}] \ \text{mod} \  \Lambda_c$ where $\b{v}$ is the lattice codeword for the private packet $\b{s}_{j,i}$.
This depends on the value of $b_i$, which is known to the user.
Let $\b{d}_1$ and $\b{d}_2$ be two mutually independent dithers that are uniformly distributed over the Voronoi region $\cV_c$. 
The dithers are known to both the user and the servers.
Then, each server transmits
\begin{equation} \label{eq:trans}
        \begin{split}
        \b{x}_j=[ \b{A}_j-\b{d}_l]  \text{ mod } \Lambda_c,
        \end{split}
\end{equation}
Where $l=1$ if $j$ is odd and $l=2$ otherwise.
\textcolor{black}{Using two dithers has several critical benefits in achieving the AWGN capacity with lattice codes, irrespective of privacy constraints.
First, it ensures $\bx_j$ and $\bx_{j+1}$ independent. Second, it allows the distribution of $\bx_j$ and $\bx_{j+1}$ to be uniform on the Voronoi region, ensuring that the input power exactly meets the power constraint for every codeword and decorrelates the estimation error from the channel input
\cite{erez2004achieving}\cite{nazer2011compute}.}
The user may add additional information to the query, informing the server which group he belongs to and which dither to use in the encoding.
This information does not affect
the privacy constraint as shown in the original paper (page 11, \cite{shmuel2021private}).

Over a transmission of $n$ symbols, the received input at the user is thus,
$$\mathbf{y}=\sum_{k=1}^{N} \mathbf{x}_{k} + \mathbf{z} = \left\lfloor \frac{N}{2}\right\rfloor\mathbf{x}_{1}+\left\lfloor \frac{N}{2}\right\rfloor\mathbf{x}_{2} + \mathbf{z}$$

In order to decode $\b{v}$, the user computes the following
\begin{equation}
\hat{\mathbf{v}}=\left[\alpha'\mathbf{y} +\mathbf{d_1}+\mathbf{d_2}\right]\text{mod} \ \Lambda_c 
\end{equation}
where $\alpha'=\alpha\frac{1}{\floor{\frac{N}{2}}}$ and $\alpha$ will be given below. The above reduces to the Modulo-Lattice Additive Noise (MLAN) channel \cite{erez2004achieving} as follows,
% \begin{fleqn}[\parindent]
\begin{equation*}
\begin{split}
\hat{\mathbf{v}}&=\left[\alpha'\mathbf{y} +\mathbf{d_1}+\mathbf{d_2}\right]\text{mod} \ \Lambda_c
% \\&=\left[\alpha \frac{1}{\left\lfloor \frac{N}{2}\right\rfloor }\mathbf{y}+\mathbf{d_1}+\mathbf{d_2}\right]\text{mod} \ \Lambda_c 
\\ &\overset{(a)}{=}  
\left[\alpha (\mathbf{x_1}+\mathbf{x_2}+\frac{1}{\left\lfloor \frac{N}{2}\right\rfloor}
\mathbf{z})+\mathbf{d_1}+\mathbf{d_2}\right]\text{mod} \ \Lambda_c 
\\ &\overset{(b)}{=}\bigg[\mathbf{x_1}+\mathbf{x_2}+(\alpha-1)(\mathbf{x_1}+\mathbf{x_2})
\\& \qquad \qquad \qquad \qquad +\alpha \frac{1}{\left\lfloor \frac{N}{2}\right\rfloor } \mathbf{z}+ \mathbf{d_1}+\mathbf{d_2}\bigg]\text{mod} \ \Lambda_c 
\\ &\overset{}{=} 
\big[\left[\mathbf{A}_{1}-\mathbf{d_1}\right]\text{mod} \ \Lambda_c+\left[\mathbf{A}_{2}-\mathbf{d_2}\right]\text{mod} \ \Lambda_c
\\& \qquad-(1-\alpha)(\mathbf{x_1}+\mathbf{x_2})+
\alpha'\mathbf{z}+ \mathbf{d_1}+\mathbf{d_2}\big]\text{mod} \ \Lambda_c 
\\ &\overset{(c)}{=} 
\left[\b{v}-(1-\alpha)(\mathbf{x_1}+\mathbf{x_2})+\alpha' \mathbf{z}\right]\text{mod} \ \Lambda_c
\\ &\overset{}{=} 
\left[\b{v}+\mathbf{z_{eq}}\right]\text{mod} \ \Lambda_c
\end{split}
\end{equation*}
% \end{fleqn}
where (a) follows from the definition of $\mathbf{x_j}$ in which all the
servers with an odd index eventually transmit $\mathbf{x_1}$ and all the rest transmits $\mathbf{x_2}$.
(b) is the MLAN equivalent channel.
(c) follows from the distributive property (\ref{eq:dist}) and due to the structure of the answers \eqref{eq:answers} where we assume that $b_i=1$. 
In case $b_i=0$, this results in a negative sign to $\b{v}$. 
Thus, since $b_i$ is known to the user, he can correct $\hat{\b{v}}$ by multiplying with $-1$.
Finally, we define the equivalent noise term
\textcolor{black}{${\mathbf{z_{eq}}\triangleq \alpha'\mathbf{z}-(1-\alpha)(\mathbf{x_1}+\mathbf{x_2})}$} where $\EX[\mathbf{z_{eq}}]=0$. 
Since $\mathbf{d_1}$, $\mathbf{d_2}$ and $\mathbf{v}$ are independent, according to the crypto lemma \cite{erez2004achieving}, $(\mathbf{x_1}+\mathbf{x_2})$ is independent of $\mathbf{v}$ as well.
Note also that for any $j$, $\mathbf{x_j}$ is uniformly distributed over
$\cV(\Lambda_c)$. 
\textcolor{black}{
Accordingly, the second moment of $\mathbf{z_{eq}}$ approaches (for $n$ large enough \cite{nazer2011compute}) to
${\sigma^{2}=\frac{1}{n}\EX\left[\norm{\mathbf{z_{eq}}}^2\right] =2(1-\alpha)^2P+\alpha^2\left\lfloor \frac{N}{2}\right\rfloor^{-2}}$ where $\alpha$ can be optimized.}
Specifically, denote by $\sigma^2_{z'}=\left\lfloor \frac{N}{2}\right\rfloor^{-2}$, we have
$\alpha_{opt}=\frac{2P}{2P+\sigma^2_{z'}}$
and the resulting optimal second moment
is $\sigma^{2}_{eq,opt}=\frac{2P\sigma^2_{z'}}{2P+\sigma^2_{z'}}$. Hence,
\begin{equation}
    R\leq \frac{1}{2}\log^+\Brack{\frac{P}{\sigma^{2}_{eq,opt}}}
    =\frac{1}{2}\log^+\left(\frac{1}{2}+\left\lfloor \frac{N}{2}\right\rfloor^{2}P\right).
\end{equation}

\end{IEEEproof}
%%%%%%%%%%%%%%%%%%
%%%%%%%%%%%%%%%%%%
\section{Corrected Proof for Lemma \ref{lemma:gap} - Gap From The MISO Sum Capacity}\label{Appendix:gapLemmaProof}
\begin{IEEEproof}
For odd $N$, 
\begin{align*}
&C_{SR}^{MISO}-R_{PIR}^J\\
&=\frac{1}{2}\log\left( 1+ N^2P\right) - \frac{1}{2}\log^+{\left(\frac{1}{2}+\floor{\frac{N}{2}}^2P\right)}\\
&\overset{(a)}{\leq}\frac{1}{2}\log\left( \frac{1+ N^2P}{\frac{1}{2}+\floor{\frac{N}{2}}^2P}\right)\\
&\overset{(b)}{\leq} \frac{1}{2}\log\left( \frac{1+ N^2P}{\frac{1}{2}+\left(\frac{N-1}{2}\right)^2P}\right)\\
&= \frac{1}{2}\log\left( 4\right)+\frac{1}{2}\log\left( \frac{1+ N^2P}{2+\left( N-1\right)^2P}\right)\\
&= 1+\frac{1}{2}\log\left( \frac{1+ N^2P}{2+\left( N-1\right)^2P}\right)\\
&\leq 2,
\end{align*}
where $(a)$ follows since $\log^+(x)\geq\log(x)$ and the last inequality follows since $\frac{1+ N^2P}{2+\left( N-1\right)^2P}<4$ for $N\geq2$ for all $P$. For even $N$, the bound in $(b)$ is not needed, and we have $C_{SR}^{MISO}-R_{PIR}^J\leq 1$.
\end{IEEEproof}
%%%%%%%%%%%%%%%%%%
%%%%%%%%%%%%%%%%%%
\section{Corrected Proof for Theorem \ref{theorem:Fading}}\label{Appendix:FadingPIRScheme}
\begin{IEEEproof}
The user, which is interested in the message $W_i$, generates a random vector $\b{b}$ of length $M$ such that each entry is either $1$ or $0$, independently and with equal probability. Then, the user divides the servers into two non-intersecting subsets, denoted as $\cS_1$ and $\cS_2$, for which he sends the query $Q_1(i)$ to each member in $\cS_1$ and $Q_2(i)$ to each member in $\cS_2$. The queries are given as follows
\begin{equation}\label{equ-queries structure PIR scheme with fading}
\begin{aligned}
&Q_{1}(i)=a_1^{-1}\b{b}, \ \ Q_{2}(i)=a_2^{-1}(-\b{b}-\b{e}_i), \ \text{ if } b_i=0\\
&Q_{1}(i)=a_1^{-1}\b{b}, \ \ Q_{2}(i)=a_2^{-1}(-\b{b}+\b{e}_i), \ \text{ if } b_i=1,
\end{aligned}
\end{equation}  
where the scalars $a_j$, $j\in\{1,2\}$ constitute the integer vector $\b{a}$.
The servers form their answers according to the received queries \eqref{equ-queries structure PIR scheme with fading}.
Again, as in the corrected scheme in Appendix \ref{Appendix:NonFadingPIRScheme}, we assume the messages are long.
Hence, the message is broken into smaller packets and sent one by one.
We focus on transmitting one of the packets.
The rest are transmitted the same way.

Specifically, for the case of $b_i=1$, we have,
\begin{equation}\label{equ-servers answer formation fading channel}
\begin{aligned}
\b{A}_{k}&=\left[\sum_{m=1}^M Q_{k,m}(i)\phi(\b{s}_{j,m})\right] \ \text{ mod }  \Lambda_c \\
&=\left[a_1^{-1}\left(\sum_{m\neq i}^M b_m \phi(\b{s}_{j,m})+\phi(\b{s}_{j,m})\right)\right] \ \text{ mod }  \Lambda_c, \\ 
\b{A}_{k+1}&=\left[\sum_{m=1}^M Q_{k+1,m}(i)\phi(\b{s}_{j,m})\right] \ \text{ mod }  \Lambda_c\\
&=\left[a_2^{-1}\left(-\sum_{m\neq i}^M b_m\phi(\b{s}_{j,m}) \right)\right]\ \text{ mod }  \Lambda_c,
\end{aligned}
\end{equation}
and for $b_i=0$ we have,
\begin{equation}
\begin{aligned}
&\b{A}_{k}=\left[a_1^{-1}\left(\sum_{m\neq i}^M b_m \phi(\b{s}_{j,m})\right)\right] \ \text{ mod }  \Lambda_c, \\ 
&\b{A}_{k+1}=\left[a_2^{-1}\left(-\sum_{m\neq i}^M b_m\phi(\b{s}_{j,m}) -\phi(\b{s}_{j,m})  \right)\right]\ \text{ mod }  \Lambda_c.
\end{aligned}
\end{equation}
Note that due to the modulo operation, the scaling does not affect the transmission power constraint.

Following quite similar steps as the proof of Theorem \ref{theorem:NonFading}, we have, 
\begin{equation*}
\begin{aligned}
& \b{x}_{j}=[ \b{A}_{l}-\b{d}_l]  \text{ mod } \Lambda_c \quad \text{for,}  \quad j\in{\cS_l},\ l=1,2. \\
\end{aligned}
\end{equation*}
Thus, over a transmission of n symbols, we get,
\begin{equation}\label{eq-channel output after subset devision}
\begin{aligned}
\b{y}&=\sum_{k\in\cS_1}h_k\b{x}_k+\sum_{k\in\cS_2}h_k\b{x}_k+\mathbf{z}\\
&=\tilde{h_1}\b{x}_1+\tilde{h_2}\b{x}_2+\mathbf{z},\\
\end{aligned}
\end{equation}
where $\tilde{\b{h}}=(\sum_{k\in\cS_1}h_k,\sum_{k\in\cS_2}h_k)$ and $\b{x}_1$ and $\b{x}_2$ are the codewords any server in $\cS_1$ and $\cS_2$ transmitted, respectively.

To decode $\b{v}$, the user computes the following
\begin{equation*}
\hat{\b{v}}=\left[\alpha\b{y}+a_1\b{d}_1+a_2\b{d}_2\right]\ \text{ mod } \Lambda_c
\end{equation*}
where $\alpha_{opt}=\frac{P\b{h}^T\frac{1}{2}\b{a}}{1+P\|\b{h}\|^2}$ (explained below), and the vector $\b{a}$ is the coefficient vector of the linear combination the user tries to decode from the noisy sum $\tilde{h_1}\b{x}_1+\tilde{h_2}\b{x}_2 + \b{z}$.
Similarly, we have the MLAN as follows,
\begin{align*}
\hat{\b{v}}&=[\alpha\b{y}+a_1\b{d}_1+a_2\b{d}_2]\text{ mod } \Lambda_c\\
&=\Big[\alpha \left(\tilde{h_1}\b{x}_1+\tilde{h_2}\b{x}_2\right)+\alpha\mathbf{z} +a_1\b{d}_1+a_2\b{d}_2\Big]\text{ mod } \Lambda_c\\
&\overset{(a)}{=}\Big[\left(\psi_1+a_1\right)\b{x}_1 +\left(\psi_2+a_2\right)\b{x}_2+\alpha\mathbf{z}\\ &\quad\quad+a_1\b{d}_1+a_2\b{d}_2)\Big]\text{ mod } \Lambda_c\\
%&=\Big[a_1\b{x}_1+a_2\b{x}_2+a_1\b{d}_1+a_2\b{d}_2\\
%&\quad\quad+\psi_1\b{x}_1 +\psi_2\b{x}_2+\alpha\mathbf{z}  \Big]\text{ mod } \Lambda_c\\
&=\Big[\left(a_1(\b{x}_1+\b{d}_1)+a_2(\b{x}_2+\b{d}_2\right)\\
&\quad\quad+\psi_1\b{x}_1 +\psi_2\b{x}_2+\alpha\mathbf{z}  \Big]\text{ mod } \Lambda_c\\
&\overset{(b)}{=}\Big[a_1\b{A}_1+a_2\b{A}_2+\psi_1\b{x}_1 +\psi_2\b{x}_2+\alpha\mathbf{z}  \Big]\text{ mod } \Lambda_c\\
&\overset{(c)}{=}\Big[\b{v}+\psi_1\b{x}_1 +\psi_2\b{x}_2+\alpha\mathbf{z}\Big]\text{ mod } \Lambda_c\\
&=\Big[\b{v}+ \b{z}_{eq}\Big]\text{ mod } \Lambda_c,
\end{align*}
where $(a)$ is by defining $\psi_l=(\alpha\tilde{h_l}-a_l)$ and $(b)$ reduces similarly to the proof of Theorem \ref{theorem:NonFading}. $(c)$ is due to the structure of the answers in \eqref{equ-servers answer formation fading channel} where we assume that $b_i=1$ and the distributive property (\ref{eq:dist}). 
In case $b_i=0$, we would result with a negative sign to $\b{v}$.
Since $b_i$ is known to the user, he can correct $\hat{\b{v}}$ accordingly.  

Here, the resulting equivalent noise is ${\b{z}_{eq} \triangleq \psi_1\b{x}_1 +\psi_2\b{x}_2+\alpha\mathbf{z}}$  and the second moment of $\b{z}_{eq}$ is approaching (for $n$ large enough \cite{nazer2011compute})
\begin{equation}
\sigma^2_{eq}=\frac{1}{n}\EX\left[\norm{\mathbf{z_{eq}}}^2\right]=\alpha^2+P\sum_{i=1}^2\psi_i^2= \alpha^2+P\Big\|\alpha\tilde{\b{h}}-\b{a}\Big\|^2.
\end{equation}
Note that $\alpha_{opt}$ is the one that minimizes the equivalent noise variance $\sigma^2_{eq}$.

Considering Theorem 2 stated in \cite{shmuel2021private}, the user is able to decode successfully and obtain $\b{v}$ (i.e., with a probability of error that tends to zero with $n$) if the lattice rate $R$ satisfies Equation (15) in \cite{shmuel2021private}. In the context of \eqref{eq-channel output after subset devision}, this results in the following, which can be further simplified as
\begin{equation}\label{equ-achievable rate for decoding the linear combination}
\begin{aligned}
R &= \frac{1}{2} \log^+ \left(\frac{1+P\|\tilde{\b{h}}\|^2}{\|\b{a}\|^2+P\left(\|\b{a}\|^2\|\tilde{\b{h}}\|^2-(\tilde{\b{h}}^T\b{a})^2\right)} \right)\\
&=\frac{1}{2} \log^+ \left(\frac{1+P\|\tilde{\b{h}}\|^2}{\sqn{a}+P\left(\sqn{a}\|\tilde{\b{h}}\|^2-(a_1\tilde{h}_1+a_2\tilde{h}_2)^2\right)} \right)\\
&=\frac{1}{2} \log^+ \left(\frac{1+P\|\tilde{\b{h}}\|^2}{\sqn{a}+P\left(a_1\tilde{h}_2-a_2\tilde{h}_1\right)^2} \right).
\end{aligned}
\end{equation}

Finally, we note that the suggested PIR scheme for the block-fading AWGN MAC does not impair the privacy requirement, and the proof is similar to the proof in \cite{shmuel2021private}.
\end{IEEEproof}
%%%%%%%%%%%%%%%%%%

%% file: figs/LatticeExample.tex
% \center
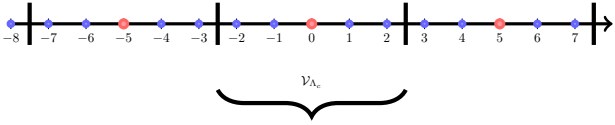
\begin{figure}\centering
\begin{tikzpicture}[scale=0.5, transform shape]
\draw[very thick,->] (-8,0) -- (8,0); % x axis
\foreach \x in {-8,-7,...,7} % numbers ticks and blue points
{
    \draw (\x cm,4pt) -- (\x cm,-4pt) node[anchor=north] {$\x$};
    \filldraw[color=blue!60, fill=blue!40, very thick](\x,0) circle (2pt);
}
\foreach \x in {-5,0,5} % red points
{
    \filldraw[color=red!60, fill=red!40, very thick](\x,0) circle (3pt);
}
\foreach \x in {-7.5,-2.5,...,7.5} % boundaries
{
    \draw[ultra thick] (\x cm,16pt) -- (\x cm,-16pt);
}
% Calligraphic brace
\draw [ultra thick,decorate,decoration = {brace,raise=25pt,amplitude=10pt,mirror}] (-2.5 cm,0) --  (2.5,0)  node[pos=0.5,below=35pt,black]{$\mathcal{V}_{\Lambda_c}$};
\end{tikzpicture}
\caption{$\Lambda_f=\mathbb{Z}$, $\Lambda_c=5\mathbb{Z}$, i.e., $\Lambda_c\subseteq\Lambda_f$.}
\label{fig:NestedLattice}
\end{figure}